\documentclass[
 reprint,
superscriptaddress,
 amsmath,amssymb,
 aps,
]{revtex4-2}

\usepackage{graphicx}
\usepackage{dcolumn}
\usepackage{bm}

\newcommand{\EE}[2]{\ensuremath{{#1}\times 10^{#2}}}

\begin{document}

\title{Quantum Phase Transitions in Nuclei and the Origin of Rhodium}

\author{P. Agarwal}
\affiliation{Department of Physics and Astronomy, Michigan State University, East Lansing, MI 48823}
\affiliation{Facility for Rare Isotope Beams, Michigan State University, East Lansing, MI 48823}

\author{H. Schatz$^*$}
\affiliation{Department of Physics and Astronomy, Michigan State University, East Lansing, MI 48823}
\affiliation{Facility for Rare Isotope Beams, Michigan State University, East Lansing, MI 48823}

\author{A. Francis}
\affiliation{Department of Physics and Astronomy, Michigan State University, East Lansing, MI 48823}
\affiliation{Facility for Rare Isotope Beams, Michigan State University, East Lansing, MI 48823}

\author{F. Montes}
\affiliation{Department of Physics and Astronomy, Michigan State University, East Lansing, MI 48823}
\affiliation{Facility for Rare Isotope Beams, Michigan State University, East Lansing, MI 48823}

\author{B. Pol}
\affiliation{Department of Physics and Astronomy, Michigan State University, East Lansing, MI 48823}
\affiliation{Facility for Rare Isotope Beams, Michigan State University, East Lansing, MI 48823}

\author{L. Roberts}
\affiliation{Computing and Artificial Intelligence, Los Alamos National Laboratory, Los Alamos, NM, USA}

\date{\today}

\begin{abstract}
The astrophysical rapid neutron capture process (r-process) is a major nucleosynthesis process responsible for the production of elements heavier than copper. Prominent features in the observationally inferred r-process abundance distributions provide critical clues for astrophysical conditions and possible r-process sites. The lack of clear features for the r-process elements lighter than tellurium has led to a broad range of proposed possible scenarios and alternative processes that may produce these elements. Here we show that the enhancement found in the solar system of rhodium and ruthenium isotopes compared to neighboring isotopes can serve as such a feature. The enhancement can be traced back to sudden changes in shape of unstable rare isotopes in the path of the light element r-process. Such shape changes have been described as quantum phase transitions. This finding points to r-process scenarios with relatively high neutron densities and temperatures for the origin of rhodium and ruthenium in the solar system. 
Rhodium and ruthenium abundances observed in r-process enhanced stars can now be used as diagnostics for the r-process conditions that produce these lighter r-process elements. Observational data for these elements indicate that different types of r-processes may have operated in the early Galaxy, and that the conditions do not necessarily align with the weak and main r-process classifications used in the past. 

\end{abstract}

\maketitle

\section{Introduction}

\begin{figure}[tbp]
\includegraphics[width=1.0\linewidth]{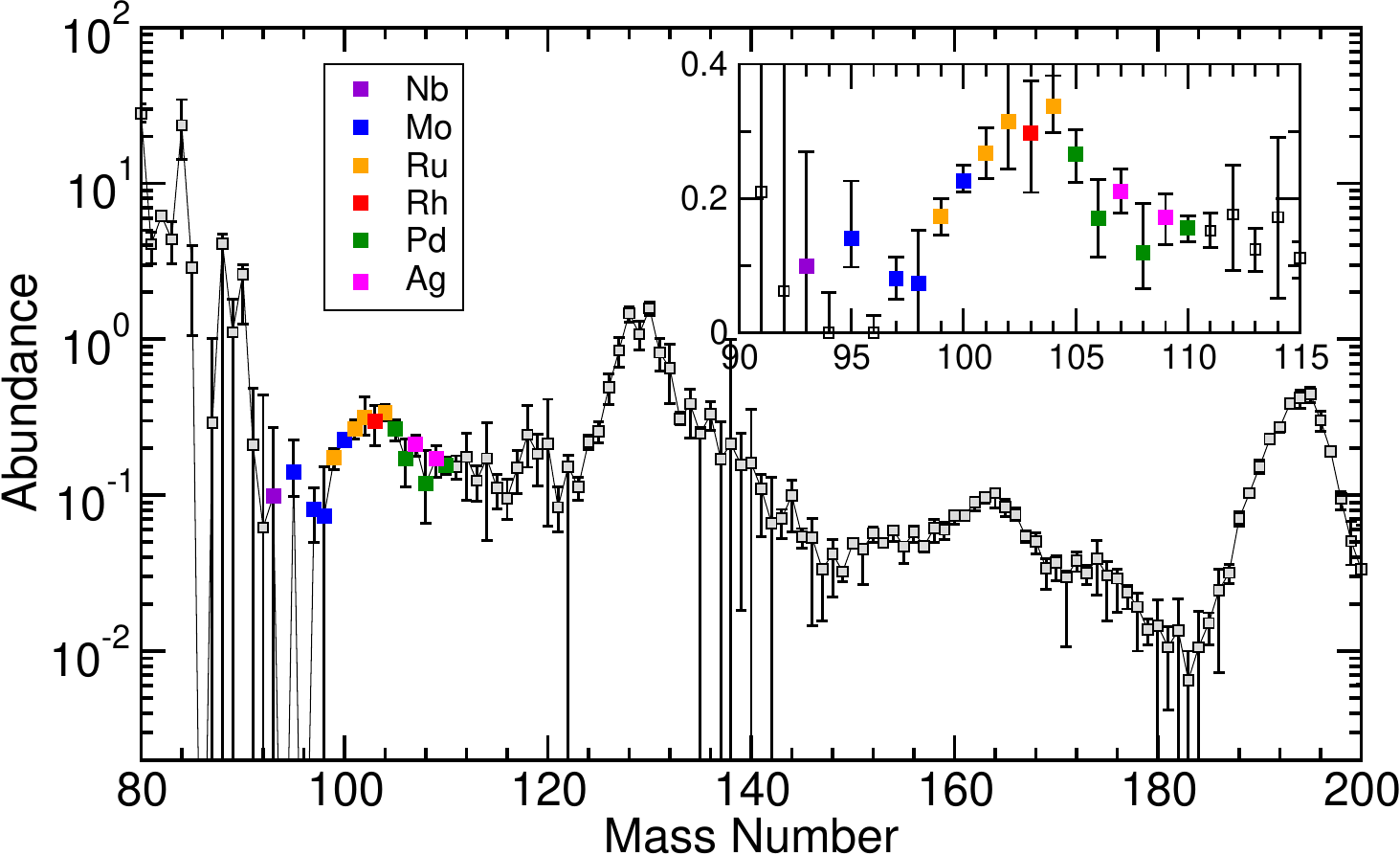}
\caption{Solar r-process abundances as functions of mass number \cite{gorielyUncertaintiesSolarSystem1999}. The inset zooms in on the ruthenium rhodium peak, and colors indicate corresponding elemental abundances in the region of interest.
\label{fig:solar}}
\end{figure}

The origin of about half of the elements heavier than copper has been attributed to the rapid neutron capture process (r-process) \cite{arnouldRprocessStellarNucleosynthesis2007, horowitzRprocessNucleosynthesisConnecting2019, cowanOriginHeaviestElements2021a, thielemannRProcessHistoryRequired2026}. 
Astrophysical site(s), conditions, and sequence of nuclear reactions of the r-process have not been identified with certainty. Possible sites include neutron star mergers, magneto-hydrodynamically driven supernova jets, collapsars, and others. 
Clues about the nature of the r-process come from features in the r-process isotopic abundance distribution in the solar system, inferred from the solar isotopic abundances by subtracting contributions from the slow neutron capture process (Fig.~\ref{fig:solar}). The most prominent features are the shoulder at mass number $A=80$, and the prominent peaks at $A=130$ and 195. These features have been linked to the classic nuclear shell closures at neutron numbers $N=50,$ $82$, and 126 
more than 65 years ago \cite{burbidgeSynthesisElementsStars1957, cameronNuclearReactionsStars1957}. The implied crossing of the neutron shell closures by the r-process reaction sequence at $A=80$, 130, and 195 directly implies a neutron capture process involving unstable neutron-rich nuclei and led to the original proposal of the existence of an r-process in the first place. The origin of the additional broader rare earth peak at $A=164$ remains a matter of debate. In 1957, \citet{burbidgeSynthesisElementsStars1957} proposed nuclear shape changes inbetween shell closures, while \citet{cameronNuclearReactionsStars1957} proposed fission of heavy nuclei beyond uranium at the end of the r-process. This controversy continues today \cite{surmanSourceRareEarthElement1997, mumpowerFormationRareearthPeak2012,gorielyNewFissionFragment2013a}. 

Here we focus on the possible nuclear physics origin of the relatively small but pronounced abundance peak at around $A=103$ (Fig.~\ref{fig:solar}). The peak is clearly identifiable within uncertainties. There are two r-only (no s-process contribution) isotopes at $A=100$ and $A=104$ that also indicate the presence of the peak, independent of uncertainties in the s-process subtraction. In terms of elemental abundances, the peak boosts by about a factor of 2-3 the abundances of ruthenium and rhodium, with the latter having a single isotope at $A=103$ in the center of the peak. 

Observations of r-process enhanced old metal poor stars that record signatures 
of nucleosynthesis events in the early Galaxy indicate multiple contributing processes to elements around the peak region, such as  co-production with heavy r-process elements and a weak r-process producing predominantly the light r-process elements (e.g., \cite{qianStellarAbundancesEarly2000, aokiSpectroscopicStudiesVery2005, montesNucleosynthesisEarlyGalaxy2007, hansenHowManyNucleosynthesis2014}). The much lower neutron-to-seed ratios required for the weak r-process and the lack of clear features in the abundance distribution have led to a broad range of proposed astrophysical sites and mechanisms. Both, neutron driven and proton or $\alpha$-particle driven processes, for example in supernovae, have been shown to reproduce observed abundances \cite{montesNucleosynthesisEarlyGalaxy2007, psaltisNeutrinodrivenOutflowsElemental2024a, frohlichNeutrinoInducedNucleosynthesisA$>$642006,farouqiNucleosynthesisModesHighEntropy2009}, including the "shoulder" at $A=80$. 

Here we address this challenge by identifying the possible nuclear physics origin of the $A=103$ rhodium peak and linking it to a neutron capture process with specific conditions. This also opens up the possibility to use stellar observations of ruthenium and rhodium to distinguish different types of nucleosynthesis scenarios such as ones with proton-rich and ones with neutron-rich conditions. 

\section{Results}
\begin{figure}[htbp]
    \centering
    \includegraphics[width=0.4\textwidth]{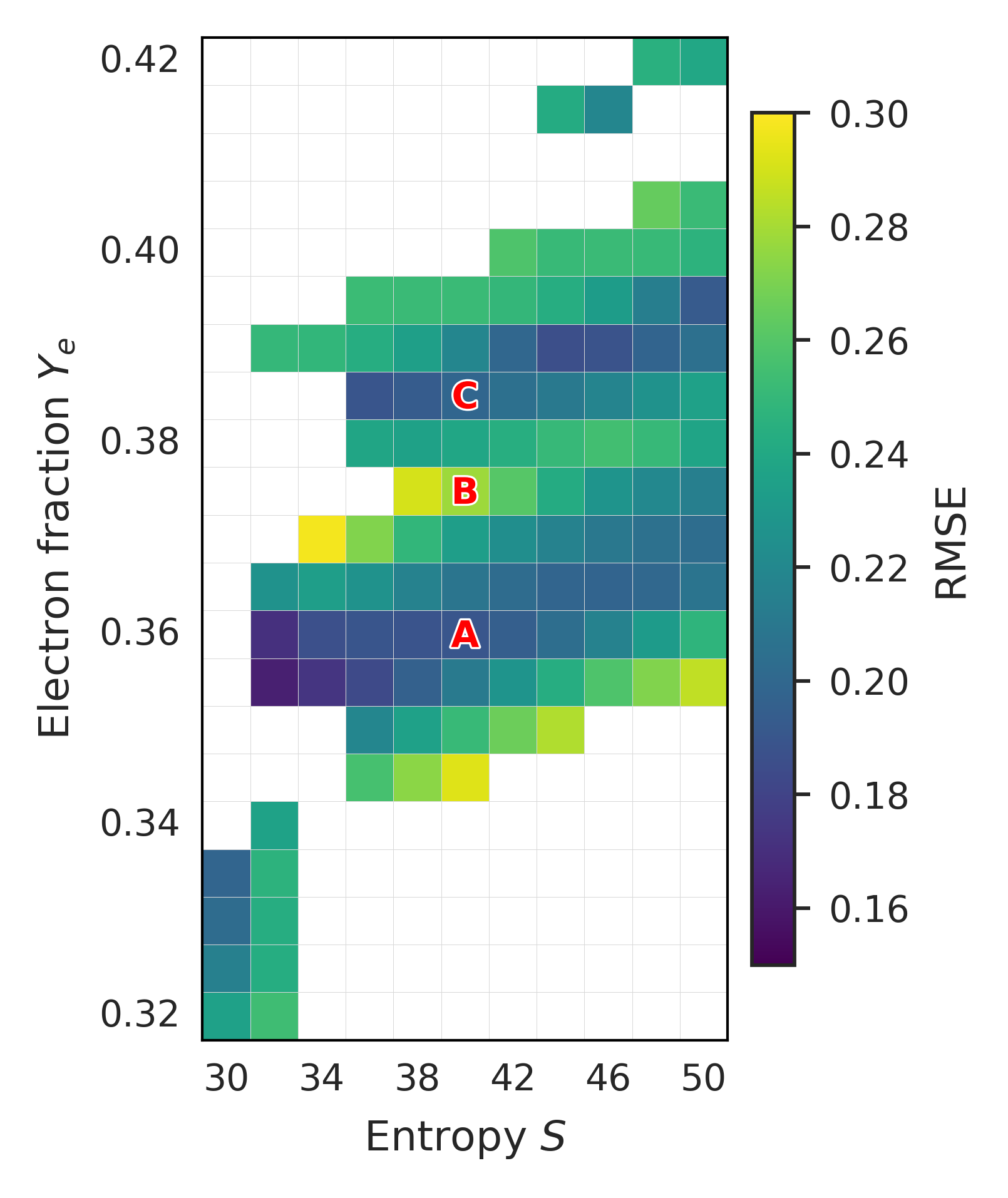}
    \caption{Root-mean-square difference between isotopic abundances predicted by the r-process model and the solar r-process abundances in the $A=103$ peak region ($A=98$--$108$) as a function of entropy and electron fraction, using the FRDM12 mass model. Indicated are also conditions A, B, and C discussed in this work.}
    \label{fig:frdm_grid}
\end{figure}

We employ a parametrized r-process model based on the reaction network SkyNet \cite{Lippuner2017}. The model includes a large network of nuclear reactions, primarily neutron captures, their inverse, and $\beta$-decay that simulates the r-process in a rapidly expanding environment such as a supernova or merging neutron stars. The parametrized approach enables us to identify typical conditions in entropy, expansion timescale, and electron fraction of the composition ($Y_e$) that reproduce the $A=103$ peak (see End Matter for details). We find a relatively narrow range of conditions required to produce the $A=103$ abundance peak (Fig. \ref{fig:frdm_grid}). These conditions are broadly consistent with previous studies that identified conditions for producing elements between the first and second r-process peak, such as areas G1 and G2 in a recent parameter survey for neutrino driven winds \cite{kuskeCompleteSurveyRprocess2025}. However, our requirement to best reproduce the $A=103$ peak without major overproduction elsewhere narrows the paramater space considerably. Solutions tend to be bands with increasing $Y_{\rm e}$ and $S$. The correlation between $Y_{\rm e}$ and $S$ is expected, as a higher entropy increases the neutron to seed ratio, which can be compensated by increasing the proton to neutron ratio and thus $Y_{\rm e}$. The best fit parameter space is split into two bands, with intermediate conditions producing a worse fit. Nuclear masses play a critical role in the calculations. While the general parameter space features appear for all investigated mass models, we focus our analysis on calculations with the FRDM12 \cite{M_ller_2016} mass model predictions as a representative case, and specifically the conditions marked as A, B, and C in Fig. \ref{fig:frdm_grid}. 

\begin{figure}[htbp]
    \centering
    \includegraphics[width=0.5\textwidth]{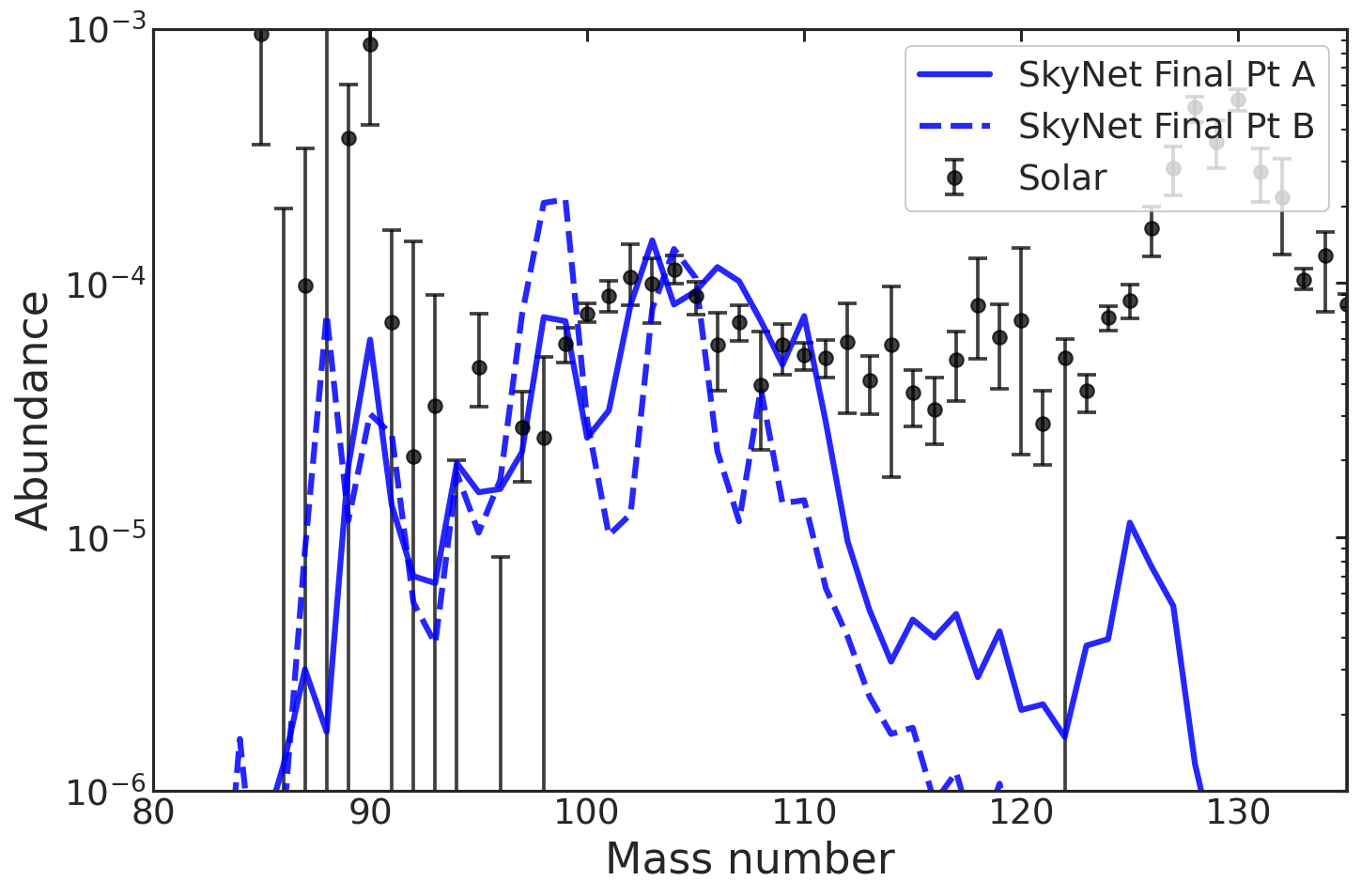}
    \caption{Final r-process model abundances as function of mass number with conditions A (solid blue) and B (dashed blue) together with solar r-process residuals (black points).}
    \label{fig:YA_PtAB_pa}
\end{figure}

\begin{figure}[htbp]
    \centering
    \includegraphics[width=0.5\textwidth]{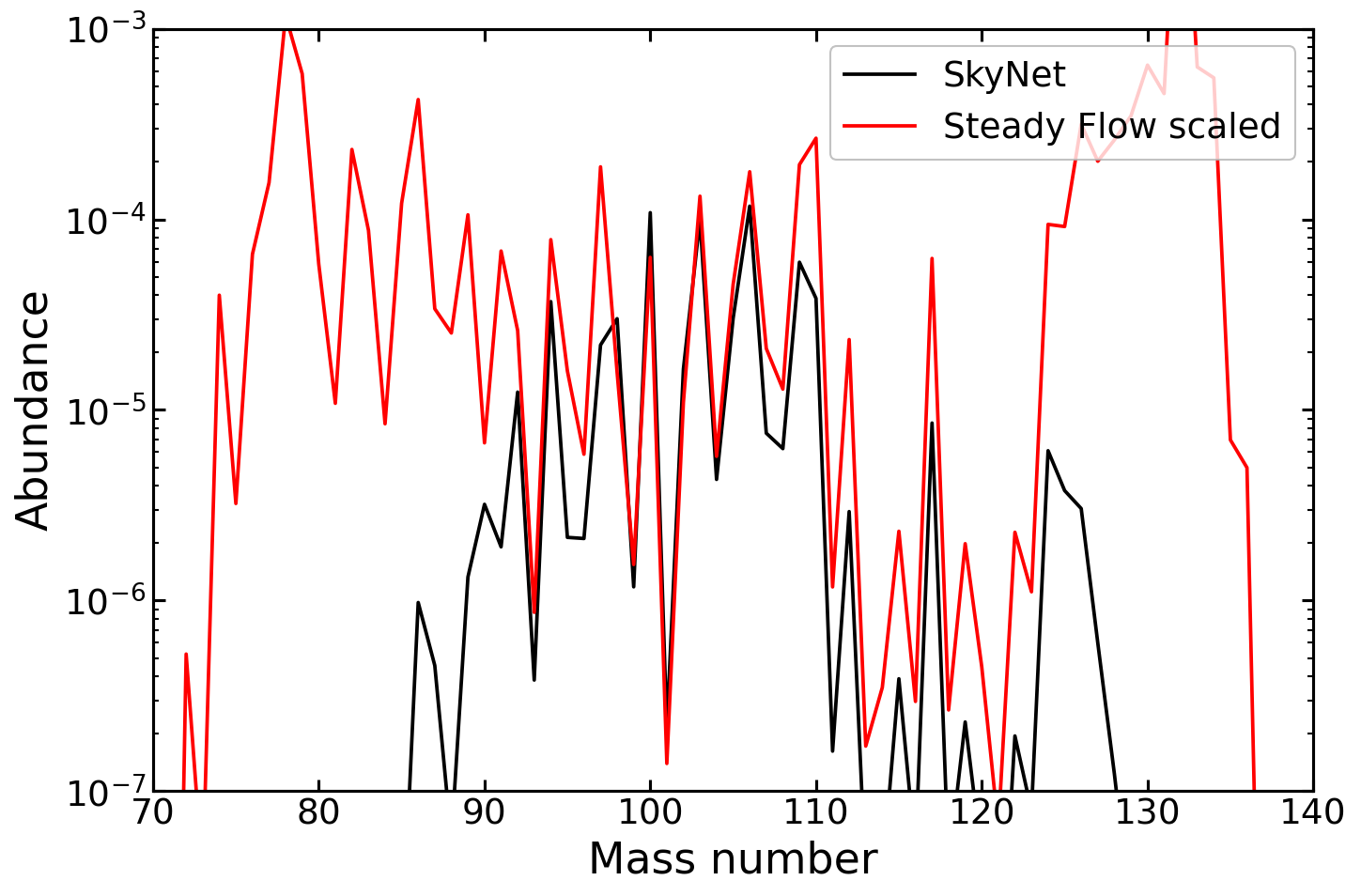}
    \caption{Dynamically calculated abundances from the r-process model after 47~ms as function of mass number (black) compared with a calculation assuming steady flow equilibrium (red) at the same temperature and neutron density conditions. }
    \label{fig:YA_PtA_pa}
\end{figure}

Conditions A reproduce the peak region reasonably well, given the nuclear uncertainties (Fig. \ref{fig:YA_PtAB_pa}). To identify the peak formation mechanism we track the evolution of the abundance distribution as a function of time and find that the mass number $A=103$ peak structure forms 47~ms after the beginning of the expansion at a temperature $T_9=1.84$ and a neutron density $n_n=\EE{2.8}{25}$~cm$^{-3}$ (Fig.~\ref{fig:YA_PtA_pa}). To disentangle nuclear structure effects from dynamic effects we compare with results from a steady flow calculation (e.g., \cite{kratzIsotopicRprocessAbundances1993}) performed at those conditions. Such a calculation assumes equilibrium between (n,$\gamma$) neutron capture and ($\gamma$,n) photo-disintegration within each isotopic element chain, and steady flow equilibrium in the $\beta$-decay from isotopic chain to isotopic chain. The resulting abundances depend entirely on neutron separation energies $S_{\rm n}$ (determined from nuclear masses) and the average decay half-lives of each isotopic chain, and are independent of time. It is well known that due to the dynamic nature of the r-process, steady flow is not typically achieved. However, abundances within specific regions between major closed shells evolve towards the steady flow equilibrium distribution. This is the basic mechanism for how nuclear structure effects imprint onto the r-process abundance pattern \cite{kratzIsotopicRprocessAbundances1993}. 

Fig.~\ref{fig:YA_PtA_pa} confirms that the dynamically calculated abundances are quite similar to ones calculated with the steady flow assumption.  In particular, the shape of the peak also appears in steady flow, with a clear reduction of abundances for $A>110$, as well as some decrease for $A<97$. For $A<90$, half-lives are too long for steady state to be achieved on the event timescale of 47~ms at this stage of the process. For $A>122$, the deviations are a combination of longer half-lives and dynamic effects due to the bottleneck responsible for the decrease around $A=110$ and the fact that the process ends before building up the $A=130$ r-process peak.

\begin{figure*}[htbp]
    \centering
    \includegraphics[width=0.7\textwidth]{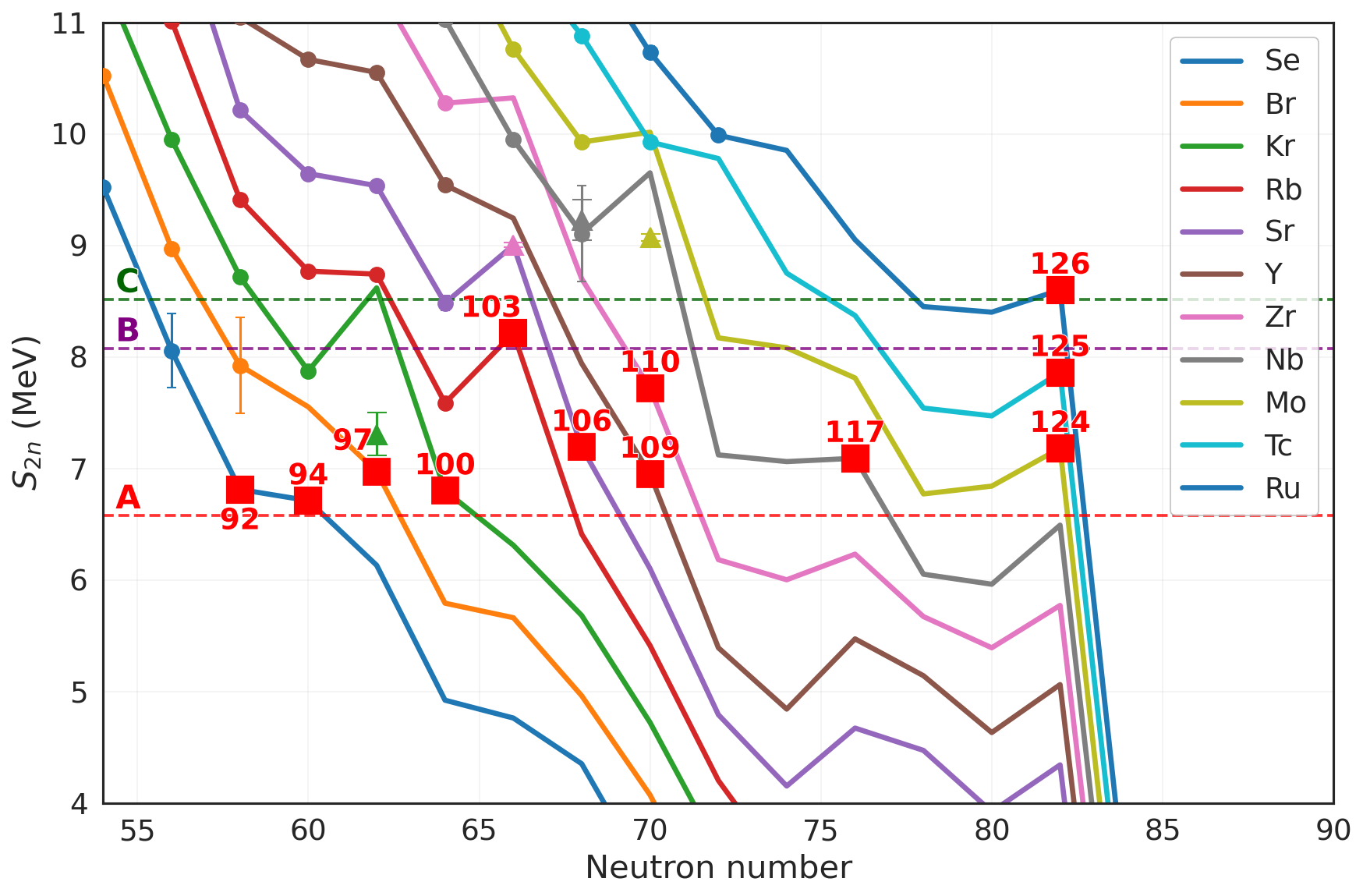}
    \caption{Two neutron separation energy ($S_{2n}$) for various isotopic element chains as function of neutron number from experiment (data points with lines) and theory (lines without data points). The equilibrium $S_{2n}$ for conditions A, B, C is indicated has horizontal dashed lines in red, purple, and teal, respectively. For condition A we also indicate the most abundant isotope in each isotopic chain together with its mass number $A$ (red squares). In $(n,\gamma)-(\gamma,n)$ equilibrium the most abundant isotopes are produced just before $S_{2n}$ drops below the equilibrium value. 
    }
    \label{fig:S2n}
\end{figure*}

Using the steady state model, we identify systematic trends in the two neutron separation energies, $S_{\rm 2n}$, and thus trends in nuclear mass, as the primary mechanism for forming the $A=103$ peak (Fig.~\ref{fig:S2n}). In (n,$\gamma$)-($\gamma$,n) equilibrium, the most abundant isotope in an elemental chain is the even $N$ isotope beyond which $S_{\rm 2n}$ drops below the  $S_{\rm 2n}$ equilibrium value for the given conditions. 
We find that the most important peak shaping feature is the sudden flattening of the $S_{\rm 2n}$ trend at $N>72$ in the Zr and Nb isotopic chains that suppresses the production of most isotopes between $A=110$ and the $N=82$ shell closure. On the low $A$ side the peak is shaped by the flattening of $S_{\rm 2n}$ around $N=60$. This leads to a broader range of isotopes close to the equilibrium $S_{\rm 2n}$ resulting in abundances distributing over a larger number of isotopes and thus lower abundances per isotope. In addition, the shift of the equilibrium towards more neutron rich nuclei helps shift abundance into the peak.

\begin{figure}[htbp]
    \centering
    \includegraphics[width=0.48\textwidth]{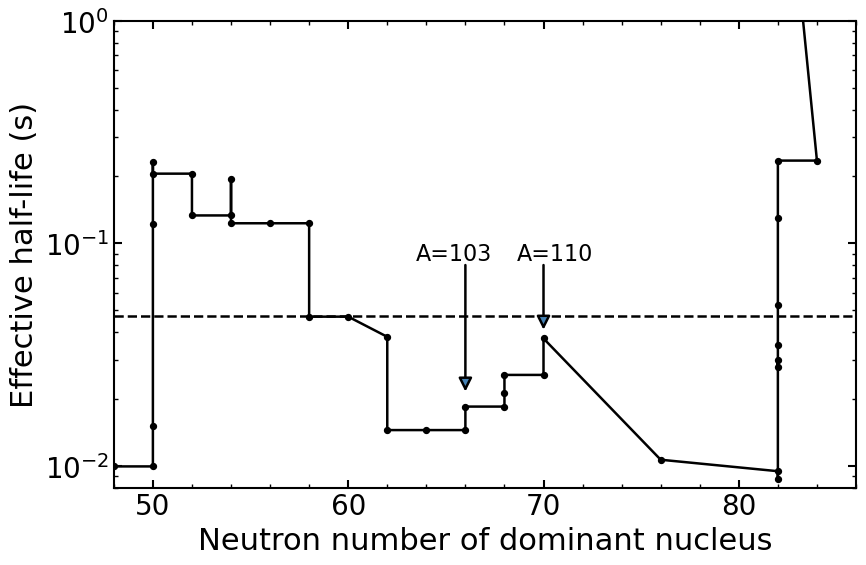}
    \caption{Effective half-life in an isotopic chain, displayed at the neutron number of the most abundant isotope in (n,$\gamma$)-($\gamma$,n) equilibrium at peak formation conditions. The data points are ordered and connected by increasing element number following the flow of the r-process. The dashed line indicates the r-process timescale at the time of peak formation. }
    \label{fig:T12}
\end{figure}

While the $S_{\rm 2n}$ select the populated isotopes, the actual abundance accumulated in such a favored isotope is also proportional to the effective average half-life in the isotopic chain. This picture breaks down for half-lives longer than the event time scale, where there is not enough time to get close to equilibrium. The $N=60$ effect in $S_{\rm 2n}$ pushes the 2n separation energy contours on the charts of nuclides to more neutron rich nuclei and thus creates a low half-life "bucket" between $N=60$ and the major shell closure at $N=82$ (Fig.~\ref{fig:T12}) where half-lives are shorter than the event timescale and abundances can get close to local steady flow equilibrium. In addition, the $N=72$ effect in $S_{\rm 2n}$ leads to a region of increased half-lives at $A=103-110$ that stay just below the event timescale, which further supports the formation of an abundance peak. Thus we find that it is the combination of $S_{\rm 2n}$ (and thus nuclear masses) and half-life systematics that shapes the $A=103$ abundance peak.

At conditions for points B and C the mechanisms are similar. Fig.~\ref{fig:S2n} indicates the preferred $S_{2n}$ values for the conditions when the peak first forms ($T_9$=1.84, $n_n=\EE{2.6}{23}$~cm$^{-3}$ for B and $T_9$=2.01, $n_n=\EE{7.0}{23}$~cm$^{-3}$ for C). For C the situation is similar to A, with the $N=60$ and $N>72$ flattening of the $S_{2n}$ trend shaping a $A=103$ peak, except that now the relevant isotopic chains are Rb for $N=60$ and Tc and Ru for $N>72$. However, for B the unusually high $S_{2n}$ for $N=62$ Kr, which is even higher than $N=60$ Kr predicted by the FRDM12 mass model, together with the experimentally determined, relatively steep drop of $S_{2n}$ beyond $N=62$ Nb, leads to both the Kr and the Nb chains to have their preferred abundance at $N=62$. This results in the overproduction at $A=98,99$ (see Fig.~\ref{fig:YA_PtAB_pa}) that disfavors conditions B over A and C and leads to the double band structure in Fig.~\ref{fig:frdm_grid}.

\section{Discussion}

The $N=60$ flattening of the $S_{2n}$ trend with neutron number for elements down to Rb (Fig.~\ref{fig:S2n}) is well established experimentally and associated with the extensively studied sudden onset of deformation at $N=60$ in the $A=100$ region of the nuclear chart \cite{heydeShapeCoexistenceAtomic2011}. This drastic change in shape with the addition of just 1-2 neutrons has been described as a quantum phase transition \cite{togashiQuantumPhaseTransition2016,gavrielovZrIsotopesRegion2022}. Theoretical studies explain the phenomenon with proton neutron interactions that lower with increasing neutron number the energy of deformation driving shell model orbitals (e.g., \cite{federmanMicroscopicStudyShape1979}). At some point these orbitals become low enough in energy to become part of the ground-state configuration and the shape changes. As Fig.~\ref{fig:S2n} shows, the FRDM12 mass model predicts a strong effect down to Kr, while for Br and Se the $S_{2n}$ slope change at $N=60$ becomes smaller. A recent mass measurement of $^{98}$Kr performed after conclusion of this study \cite{lunneyExtendingLowZBorder2025} indicates, that this more gentle behavior already sets in at Kr. This is in line with Coulomb excitation and $\gamma$-spectroscopy experiments indicating in the Kr isotopic chain a more gradual onset of deformation at $N=60$ \cite{albersEvidenceSmoothOnset2012,flavignyShapeEvolutionNeutronRich2017}. This likely alleviates the overproduction issue for condition B and may reduce the 2 band structure of the favorable parameter space in Fig.~\ref{fig:frdm_grid}. Nevertheless, a moderate flattening of the $S_{2n}$ trend is still present, supporting the formation of the $A=103$ abundance peak. Whether this effect indeed continues down to Se remains to be demonstrated experimentally. 

The more extensive flattening of the $S_{2n}$ systematics beyond $N=72$ in the here relevant Zr-Rh element range is predicted by most mass models. It is again related to rapid shape changes from the strong axially prolate deformed nuclei beyond $N=60$ to spherical or oblate shapes as predicted by the FRDM12 and many mean field type theoretical approaches \cite{stoitsovSystematicStudyDeformed2003, miyaharaShapeEvolutionZr2018,	blazkiewiczCoordinateSpaceHartreeFockBogoliubov2005}. Experimentally, strong prolate deformation in the Zr and Mo isotopic chains, with some possible admixture of tri-axial shapes, has been confirmed via $\gamma$-ray spectroscopy out to $N=70$ \cite{paulAreThereSignatures2017, moonTriaxialDeformationNeutronrich2024a, moonFirstHighresolutionInbeam2025}. However, no deformation data are available beyond $N=70$. $\beta$-decay half-lives could in principle be used to constrain nuclear shapes in this region \cite{yoshidaDecayHalflivesIndicator2023}. Indeed, in the Mo isotopic chain, $\beta$-decay half-lives have been measured out to $N=76$ $^{118}$Mo \cite{lorusso$ensuremathbeta$DecayHalfLives1102015}. Interestingly, the experimental data show a significantly different trend compared to FRDM12 based predictions, but the data have not been analyzed yet in terms of deformation.

\section{Conclusions}
With the identification of a mechanism for the production of an $A=103$ Rh peak in the r-process, the presence, or absence, of such a peak can now be used as diagnostics to constrain r-process site conditions. The presence of a peak in the solar r-process abundance distribution indicates that the non s-process Rh and Ru in the solar system originate from an r-process with relatively high neutron densities that achieved (n,$\gamma$)-($\gamma$,n) equilibrium and passed through the region of neutron rich nuclei with rapid shape changes at neutron numbers $N=60$ and $N=72$. Proton capture processes such as the $\nu$p-process, charged particle processes, or weak r-processes with low neutron densities are unlikely to be main contributors. 

\begin{figure}[htbp]
    \centering
    \includegraphics[width=0.48\textwidth]{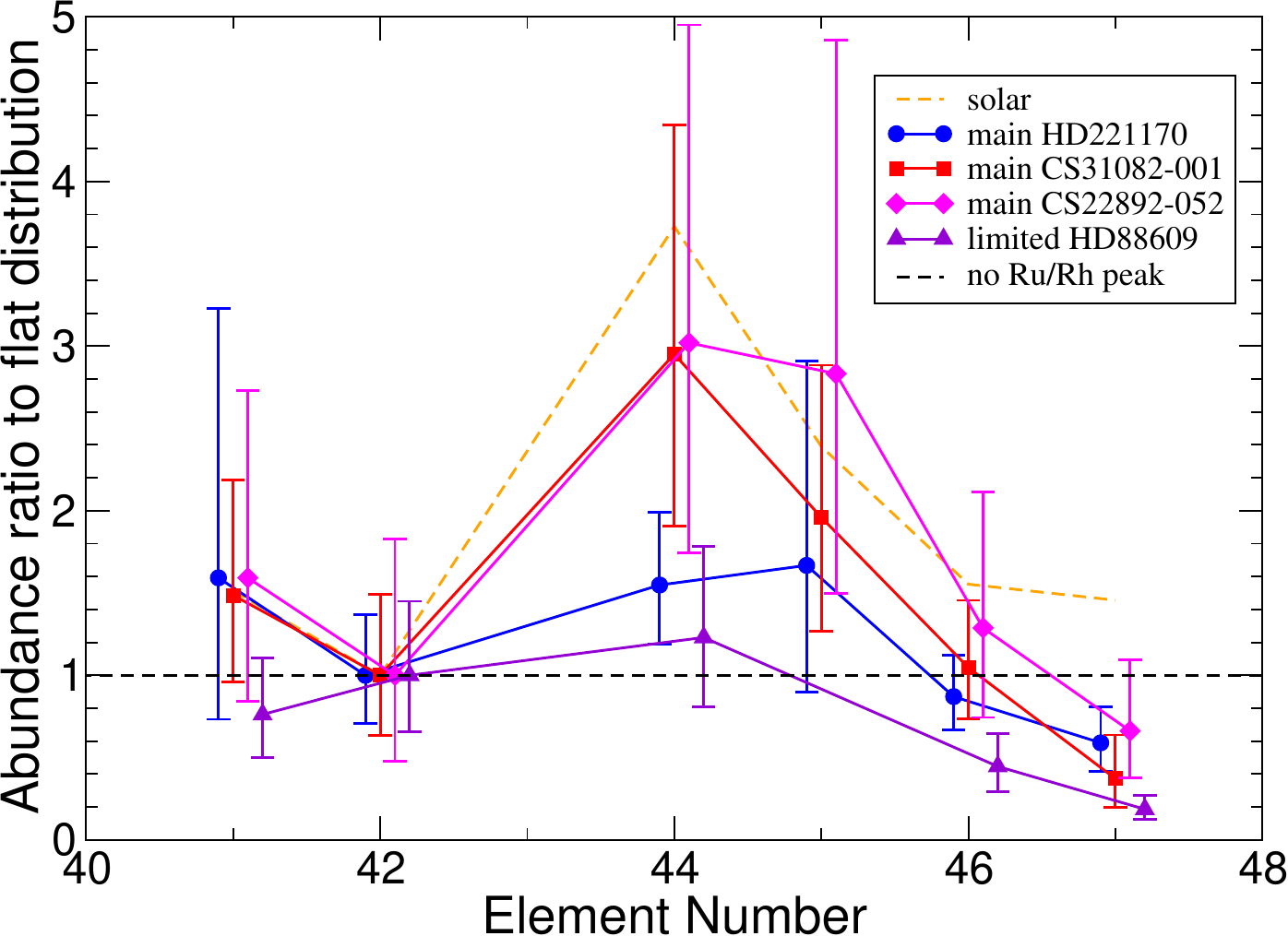}
    \caption{Ratio of elemental abundances to expected elemental abundances for a flat isotopic abundance distribution for the solar system r-process abundances (orange dashed line) and 5 metal poor stars: HD221170 \cite{ivansNearUltravioletObservationsHD2006} (blue circles), CS31082-001 \cite{siqueiramelloFirstStarsXVI2013} (red squares), CS22892-052 \cite{snedenExtremelyMetalpoorNeutron2003} (magenta diamonds), and HD88609 \cite{hondaNeutronCaptureElementsVery2007} (purple triangle). The flat distribution is determined from the Mo abundance.}
    \label{fig:abundances}
\end{figure}

Beyond the solar system, Ru and Rh abundances in r-process enhanced metal poor stars can be examined. These stars record signatures of r-process events in the early Galaxy. There is clear evidence for at least two types of events and hence two classes of r-process enhanced stars \cite{qianStellarAbundancesEarly2000, aokiSpectroscopicStudiesVery2005, montesNucleosynthesisEarlyGalaxy2007, hansenHowManyNucleosynthesis2014}: a main r-process that produces all r-process elements up to the actinides (main r-process stars), and some other process, possibly a weak r-process, that produces only the light r-process elements up to around Ag (limited-r stars). However, observational uncertainties are large, and available data for Ru and Rh are limited, especially for limited-r stars (e.g., \cite{xylakis-dornbuschRProcessAllianceAnalysis2024}). Nevertheless we selected a few representative stars and determined the ratio of the observed elemental abundances to the elemental abundances that would be expected for a flat isotopic abundance distribution without a Rh peak (Fig.~\ref{fig:abundances}). The flat isotopic abundance distribution is determined from the observed Mo abundance, which samples 4 isotopes just below the peak. Typical main r-process stars such as CS31082-001 \cite{siqueiramelloFirstStarsXVI2013} and CS22892-052 \cite{snedenExtremelyMetalpoorNeutron2003} exhibit an abundance pattern that agrees very well with the solar r-process for the heavy r-process elements, but also tracks relatively closely the solar pattern for the lighter r-process elements, though deviations can be larger there \cite{cowanOriginHeaviestElements2021a}. Indeed, the Nb ($Z=41$) to Ag ($Z=47$) abundance ratios in  Fig.~\ref{fig:abundances} for CS31082-001 and CS22892-052 show the Ru/Rh peak enhancement also found in the solar system. Based on this work this indicates a high neutron density r-process producing the Ru and Rh in these stars. This is not unexpected, as a high neutron density r-process is required to produce the heavy r-process elements with the $A=130$ and $A=195$ abundance peaks also observed in these stars. However, the limited-r star HD88609 \cite{hondaNeutronCaptureElementsVery2007} has abundance ratios that are incompatible with the Rh/Ru peak and are instead consistent with a flat isotopic abundance distribution. This may indicate that the process that produces the elemental abundance signatures in limited-r stars is not a high neutron density r-process. Surprisingly, the main r-process star HD221170 \cite{ivansNearUltravioletObservationsHD2006} also appears to better agree with a flat isotopic abundance distribution. This suggests that different types of processes are responsible for the light r-process elements in different main r-process stars. A more systematic analysis of a larger sample of stars, taking into account consistency and correlations of observational uncertainties, would be desirable to confirm these initial findings. 

To map out in more detail the peak forming mechanism identified in this work, and to more tightly constrain r-process conditions, additional experimental data on neutron rich nuclei across neutron numbers $N=60$ and $N=72$ would be important. For $N=60$, mass measurements out to $N=62$ for Se and Br are needed to fully map the region of rapid shape change.  No experimental information is available for the $N=72$ shape transition in the Zr-Tc elemental isotopic chains. It would be important to perform decay studies, $\gamma$-spectroscopy studies, and mass measurements out to $N=74$ to confirm the existence of the effects and map their extent on the nuclear chart, which would directly map onto the required r-process condition space. 

\section{Acknowledgements}
This work has been supported by NSF grants PHY-2514797, PHY-2209429, and PHY-1913554. The work strongly benefited from activities supported through NSF award OISE-1927130 (International Research Network for Nuclear Astrophysics), and DOE awards DE-SC0023128 and DE-SC0026204 (CeNAM). This document is approved for unlimited release under LA-UR-26-26194.

\section{Author Contributions}
H.S. initiated the study; P.A., H.S., A.F., B.P., and R.L. wrote code and performed calculations, P.A., H.S., A.F., B.P. performed r-process analysis, F.M. contributed to analysis and interpretation of stellar data, H.S. and P.A. wrote the first draft, all authors edited the manuscript.


\section{End Matter}
We employ a parametrized r-process model based on the reaction network SkyNet \cite{Lippuner2017}. The model computes the isotopic abundance evolution, and the thermodynamical evolution taking into account nuclear heating. Neutron capture and $\beta$-decay rates were taken from JINA REACLIB \cite{Cyburt_2010}, updated with experimental $\beta$-decay rates and neutron emission branchings from NUBASE2020 \cite{Kondev_2021}. ($\gamma$,n) photodisintegration rates are calculated from neutron capture rates via detailed balance, using masses from AME20 \cite{Huang_2021}, augmented with data from the FRDM12\cite{M_ller_2016} mass model. 
For comparison, we also performed calculations using the UNEDF1\cite{Kortelainen_2012}, DZ31\cite{PhysRevC.52.R23}, and WS4\cite{WANG2014215} mass models. For these calculations, theoretical $\beta$-decay half-lives and theoretical neutron capture rates were scaled to account for the change in Q-value. $\beta$-decay rates were scaled with $Q^5$, while the scaling for neutron capture rates was estimated with the computationally fast statistical model code SMOKER \cite{thielemannThermonuclearReactionRates1986}. 
To make the calculations less computationally expensive, the network was reduced in size and limited to nuclides with proton number $Z \le 60$.

\begin{figure}[htbp]
    \centering
    \includegraphics[width=0.5\textwidth]{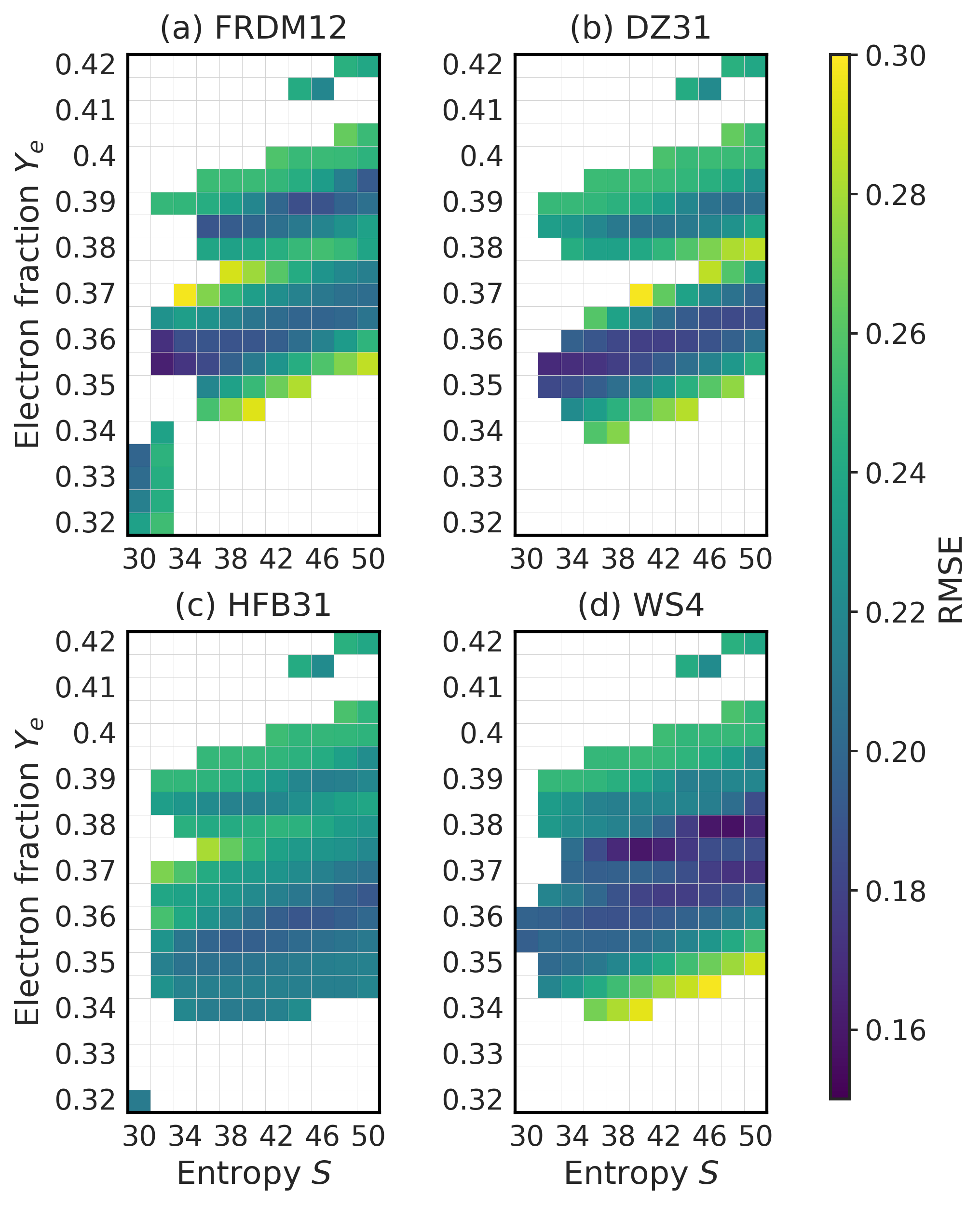}
    \hspace*{-1cm}
    \caption{Root-mean-square difference between isotopic abundances predicted by the r-process model and the solar r-process abundances in the $A=103$ peak region ($A=98$--$108$) as a function of entropy and electron fraction, using the FRDM12 mass model (a), the DZ31 mass model (b), the HFB31 mass model (c) and the WS4 mass model (d).}
    \label{fig:four_models_grid}
\end{figure}

The r-process conditions in Skynet are parametrized by initial temperature, entropy per baryon $S$, electron fraction $Y_{\rm e}$, and expansion timescale $\tau$. 
To explore the possible astrophysical environments forming an $A=103$ abundance peak, we conducted a parameter space survey sampling parameter ranges of $Y_{\rm e}$=0.2--0.45 and $S=$10--52 ($k_{\rm B}$/baryon). We chose an initial temperature of 10~GK ($T_9=10$) to ensure all calculations start in nuclear statistical equilibrium. These parameter ranges were chosen based on typical values found in models of realistic astrophysical environments such as various outflows in neutron star mergers \cite{Just_2015,Wu_2016,Radice_2018,Kullmann_2021} and jet driven supernovae \cite{Nishimura_2015}. We considered a $\tau$-range of 1-12~ms to be reasonable, but test calculations indicated no major qualitative changes over that range, hence we fixed $\tau=6$~ms. While realistic r-process environments may extend the $Y_e$ range to much lower values, those low $Y_e$ will not contribute to the light r-process elements of interest for this study. 

To evaluate the ability of a calculation to reproduce the abundance pattern around the $A=103$ abundance peak, we scaled the resulting abundance pattern $y_{\rm calc}$ to minimize the RMS residual relative
to the solar r-process abundances, $y_{\rm sol}$, over the mass number range $A=100$--105 so that $y_{\rm pred}=y_{\rm calc} \sum _A y_{\mathrm{calc}}(A)\,y_{\mathrm{sol}}(A)/\sum_A y_{\mathrm{calc}}(A)^2$. While the goal was to normalize to the maximum of the peak, we chose a narrow range around the peak to minimize dependence on isotope to isotope scatter due to nuclear physics uncertainties in the model. 
To quantify the quality of the fit we use a root mean square (rms) deviation over the
$A=98$--108 range that encompasses the entire shape of the peak. An rms was chosen as the lack of statistical scatter of the solar abundances in relation to the error bars indicates significantly correlated systematic errors. In addition, we filter out any calculations that after scaling lead to significant overproduction anywhere else. Such conditions cannot contribute significantly to the region of interest even if they match the local pattern very well. We therefore calculate from individual overproduction ratios $r(A)=\log (y_{\rm pred}(A)/y_{\rm sol}(A))$ an overproduction measure 
$\alpha=\sum_A r(A)_{>0}/n$ where we only sum positive $r(A)$ over the $n$ mass numbers comprising the non-zero $A$-range of the abundance pattern.  
As there are significant model and nuclear physics uncertainties, the overproduction criterion had to be chosen with some reasonable level of tolerance. We found that $\alpha < 0.025$ provides a reasonable cutoff. Results for r-process model calculations with the four different nuclear mass models are shown in Fig.~\ref{fig:four_models_grid}. While we performed calculations down to entropies of 10~$k_{\rm B}$/baryon we find that for entropies below 30~$k_{\rm B}$/baryon and corresponding low $Y_e$ there is significant overproduction in the $A=92-97$ mass range that exceeds the height of the $A=103$ peak. These conditions are therefore not suitable to form the $A=103$ peak. 

For comparison we also use a simple steady flow model. In steady flow, we assume (n,$\gamma$)-($\gamma$,n) equilibrium within isotopic chains, which establishes an abundance pattern that favors 1-2 even $N$ isotopes closest to a specific 2-neutron separation energy
\begin{equation}
S_{\rm 2n}=2 \frac{T_9}{11.604} \left[ \frac{3}{2} \ln(T_9)-ln(n_n)+78.460 \right] \label{EqSn}
\end{equation}
with $S_n$ in MeV and $n_n$ in CGS units. The abundance weighted $\beta$-decay lifetimes of these isotopes result in an effective $\beta$-decay lifetime $\tau_\beta(Z)$ of the isotopic chain. After a sufficiently long time, the r-process will have established in addition a steady flow equilibrium fixing the relative abundance pattern from isotopic chain to isotopic chain as $Y(Z) \tau_\beta(Z)={\rm const}$. The final abundance pattern is not time dependent anymore, and entirely determined by nuclear structure effects. However, for the composition to approach local steady flow requires $\tau_\beta(Z) < \tau$ with $\tau$ being the event timescale that can be estimated from the elapsed time when the composition is considered. To account for the reduction in abundance due to the slower buildup via feeding from slower isotopic chains, we apply to the steady flow abundances $Y(Z)$ a correction factor $[1-\exp(-\tau/\tau_\beta(Z-1))]$. We analyze the steady flow calculation using the conditions at the time of peak formation. This is the time when in the full network calculation a maximum in the peak area, and a turning point on the high A side begin to appear. As this is a gradual process, there is a range of possible choices of conditions. For example, for point A, there is a temperature range of 1.83-2~GK and a neutron density range \EE{1.9}{26}--\EE{3.3}{25}~cm$^{-3}$ over which the peak forms. We find that this uncertainty does not impact the analysis, as the corresponding equilibrium neutron separation $S_n$ only varies between 3.27--3.25~MeV.

\bibliography{110peak_2.bib,citations_pa}

\end{document}